# Machine learning for knowledge acquisition and accelerated inverse-design for non-Hermitian systems

W. W. Ahmed[1], M. Farhat[1], K. Staliunas[2,3,4], X. Zhang[1,5†], and Y. Wu[1*]

[1]*Division of Computer, Electrical and Mathematical Sciences and Engineering, King Abdullah University of Science and Technology (KAUST), Thuwal, 23955-6900, Saudi Arabia*

[2]*Departament de Física, Universitat Politècnica de Catalunya (UPC), Colom 11, E-08222 Terrassa, Barcelona, Spain*

[3]*Institució Catalana de Recerca i Estudis Avançats (ICREA), Passeig Lluís Companys 23, E-08010, Barcelona, Spain*

[4]*Vilnius University, Laser Research Center, Saulėtekio al. 10, Vilnius, Lithuania*

[5]*Department of Computer Science and Engineering, University of Notre Dame, Notre Dame, IN 46556, United States of America*

Emails: [*]ying.wu@kaust.edu.sa, [†] xzhang33@nd.edu

**Abstract:** Non-Hermitian systems offer new platforms for unusual physical properties that can be flexibly manipulated by '*redistribution*' of the real and imaginary parts of refractive indices, whose presence breaks conventional wave propagation symmetries, leading to asymmetric reflection and symmetric transmission with respect to the wave propagation direction. Here, we use supervised and unsupervised learning techniques for knowledge acquisition in non-Hermitian systems which accelerate the inverse design process. In particular, we construct a deep learning model that relates the transmission and asymmetric reflection in non-conservative settings and propose sub-manifold learning to recognize non-Hermitian features from transmission spectra. The developed deep learning framework determines the feasibility of a desired spectral response for a given structure and uncovers the role of effective gain-loss parameters to tailor the spectral response. These findings pave the way for intelligent inverse design and shape our understanding of physical reality in general non-Hermitian systems.

**Introduction:** Non-Hermitian Systems, first studied in quantum mechanics[1,2], have been attracting a growing interest in wave physics, especially in optics for being a promising platform to explore new physical phenomena[3-7] that are impossible in Hermitian systems. Hermitian systems exhibit symmetric transmission and reflection with respect to the direction of the incident wave. This symmetry stems from reciprocity and energy conservation principles. Yet, gain-loss in non-Hermitian systems breaks space symmetry and reveals unusual properties such as unidirectional invisibility[8,9], lasing and coherent perfect absorption[10], asymmetric chirality[11] and many others[12-16]. In simple cases, even if the scatterers are elastic (conservative) and only one dimensional (1D), scattering problems may not be solved analytically. Some general relations may be derived. For instance, the reflection and the transmission for conservative systems obey $|t|^2 + |r|^2 = 1$. For non-conservative but balanced gain-loss systems, another general relation exists, i.e., $r_L r_R^* = 1 - |t|^2$ relating the left and the right reflection waves. However, in the most general case, such relations do not exist. Lowest order Born approximation valid for weak scatterers predicts the left and right reflection and transmission spectra as $r_{L,R}^0(k) = \frac{ik}{2} \int_a^b \varepsilon(x) e^{\pm 2ikx} dx$ and $t^0(k) = 1 + \frac{ik}{2} \int_a^b \varepsilon(x) dx$, where for $x < a$ and $x > b$ the medium is air. These integrals depend on a general function of scatterer, $\varepsilon(x)$, and value of reflection, $r_{L,R}^0$, for left and right incident plane wave $e^{\pm ikz}$. Obviously for real valued $\varepsilon(x)$, the above relation gives $r_L = r_R^*$. Yet, continuing to calculate higher order born approximations, reflection-transmission spectra become interdependent due to recursive relation of reradiated electric fields i.e., $E_n(\zeta) = \frac{ik}{2} \int_a^b \varepsilon(x) E_{n-1}(z) e^{-ik|x-\zeta|} dz$ where $\zeta$ lies in the range of $a \leq \zeta \leq b$. This implies that the transmission and the reflection spectra might be related even for arbitrary scattering functions $\varepsilon(x)$. The analysis of Born approximation chain reveals that this hypothetic dependence is absent for very weak scattering and vanishes proportionally to $|\varepsilon(z)|^2$. Nevertheless, the analysis, even in this 1D case, cannot produce analytical tractable results or general relation between $r_L, r_R$, and $t$ spectra. Eventually, whether $r_L, r_R$ and $t$ are mutually related remains an open question. In addition, the amount of effective gain-loss plays a crucial role in designing the specific functionalities of the non-Hermitian structures that require careful tuning of design parameters. The existing modeling methods in 'conventional photonics' use exhaustive tuning of material (and geometrical) parameters via brute-force and optimization for '*on demand*' wave control[17,18], which are computationally expensive. Therefore, intelligent models to understand the underlying physics of wave-matter interaction with the latent data structures is desired to reveal the relation between different physical quantities and automate the design of technological devices.

In recent years, machine-learning (ML) techniques were successfully applied for forward and inverse designs in different physical settings[19-34]. ML based models are often regarded as '*black boxes*' because they do not reveal the physical behavior of the designed structures. In the literature, less efforts have been devoted to get scientific insights of photonic structures using ML[35-38]. Such scientific knowledge is essential to exclude suboptimal designs for fabrication or to discover new physics. Although tremendous progresses in ML-enabled methods have been made to solve different physical problems, most of the reported studies are devoted to address the Hermitian (or conservative) systems, leaving ML models for non-Hermitian problems yet to be developed. Here, we focus on non-Hermitian systems and exploit the ML for knowledge extraction that streamlines the inverse design process. By analyzing large amounts of spectral data, the ML provides us hint to the answer to the general question about the relation between asymmetric reflections and transmission in non-conservative systems. We develop powerful design tools that intelligently learn the bidirectional mapping between non-Hermitian material parameters and scattering spectra, considerably reducing the computation requirements. The forward process from design material parameters to response-space is well defined, but the backward inverse problem (from response space to material parameters) is ambiguous due to the degenerate solution space[39]. In fact, one identical response can be yield by various parameter sets. Therefore, solving the inverse design problem for the desired spectral properties is very challenging for practical realizations due to the existence of non-unique solutions. To this end, we first identify the sub-manifolds of non-Hermitian features (gain, loss, balanced gain-loss cases) which play an important role for inverse design. We use unsupervised learning based on Principal Component Analysis[40] (PCA) that reduces the dimensionality of the transmission spectra over a set of training data and discover the clustering among the data representing gain, lossy, and balanced non-Hermitian cases, as depicted in Fig.1a. The largest variance component helps to understand the underlying features in the data and provides insights about the role of effective media in designing non-Hermitian structures. Further, sub-manifold learning is utilized to accelerate the inverse design process. The inverse design process is divided into training and inference phases as shown in Fig. 1b. The training phase involves the modeling of the forward neural network, $\hat{f}(D)$ as a function of the design space, $D$, where $\hat{f}$ is a universal approximator function for forward simulator. Inference phase retrieves the design parameters, $D$, for on-demand spectra, which involves the computation of feasible starting design parameters based on manifold learning and the gradient of the pretrained forward model with respect to the design space $\partial \hat{f}/\partial D$ while fixing all the weights and biases of the network. KD-tree algorithm[41] is applied to find the nearest neighboring point for the latent transmission spectra within a feasible region that works as a promising starting point, $D$, to achieve the optimal design using adaptive gradient approach. In addition, we reconstruct the transmission spectra from the asymmetric reflection that uncovers one-to-one and one-to-many mapping among reflection and transmission in forward and reverse directions. As a proof of concept and without loss of generality, we apply our design strategy to study multilayered non-Hermitian structures where the complex optical materials exhibit highly asymmetric optical response due to simultaneous index and gain-loss modulation. This study thus contributes on developing deep learning frameworks for *'knowledge acquisition'* rather than merely *'optimization'*, which ultimately uncover the relationship between the transmission and the asymmetric reflection, recognize the non-Hermitian features in spectral data, and solve the inverse problem in general non-Hermitian systems.

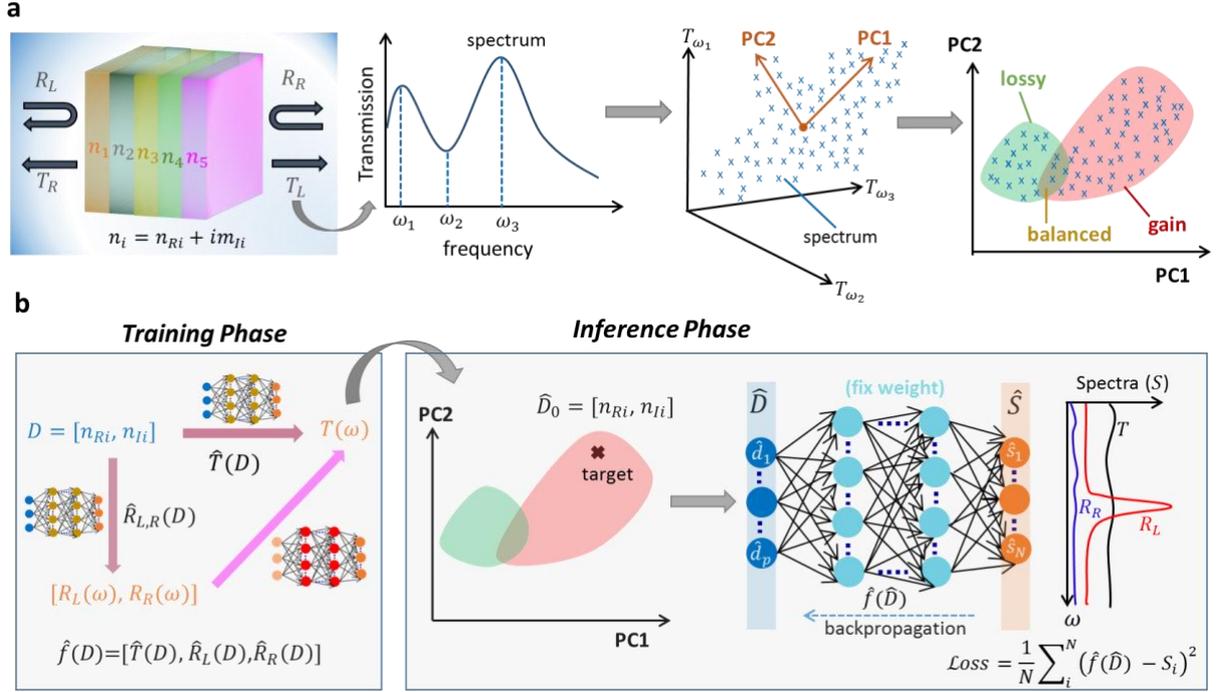

Fig. 1 Conceptual illustration of the design process for machine learning assisted non-Hermitian photonics structures. a Dimensionality reduction based on Principal Component Analysis (PCA) reveals the lower-dimensional sub-spaces where gain, lossy, and balanced system reside. A high dimensional transmission spectrum generated from a periodic five-layer photonic structure, whose unit-cell is shown, is mapped to a point. The latent space for different spectra form feasible regions of gain, lossy, and balanced systems spanned by the first two principal components (PC1 and PC2). b For the inverse design, the dimensionality of the desired response (transmission) is reduced using PCA to observe the feasibility of the response with effective gain, lossy, and balanced media. Using a trained neural network (NN) that relates the design space with the response space, we search for the optimum solution with adaptive gradient descent method. The nearest neighbor of the target response within the possible region is used as a starting point to optimize the final structure.

**Results and Discussion:** To demonstrate the effectiveness of our design approach, we study a non-Hermitian structure with periodically distributed five-layer supercells, where each layer of the unit cell may be parametrized with thickness and material as shown Fig. 1. We model such structure with transfer matrix method (TMM) where the material properties of the $i$th layer are represented by complex refractive index $n_i = n_{Ri} + in_{Ii}$, where $n_R$ and $n_I$ are the real and the imaginary parts of the refractive index [see Method section for detail]. The background medium is air. The complex refractive index distribution of the supercell dictates its scattering properties. Data driven approaches generally require a huge amount of data (from thousands to millions of sets) for discovering the hidden features and the intrinsic relations between the input and the output. In the data generation process, we assume the same thickness for each layer and consider material parameters as design variables $\mathcal{D} = [n_{R1}, n_{R2}, n_{R3}, n_{R4}, n_{R5}, n_{I1}, n_{I2}, n_{I3}, n_{I4}, n_{I5}]$ to determine the scattering response, i.e., transmission and reflection properties for the left and right incident waves. Because of reciprocity, such a structure provides symmetric transmission $T = T_R = T_L$ and, in general (e.g., if mirror symmetry is broken), asymmetric reflection denoted by $R_R$ and $R_L$ for left and right direction of incident waves, respectively. For illustration, the spectral response of interest is set in the normalized frequency range $0.2 \leq \omega a/2\pi c \leq 0.8$ where each data set contains ten design parameters and 100 discrete points for each transmission and reflection spectrum $S = [s_1, s_2, s_3, s_4 \ldots \ldots s_{100}]$. The real and imaginary parts of the refractive index are restricted in the range [1 1.4] and [-0.2 0.2] for data generation, respectively. In our study, we randomly generate 50000 data samples of forward simulations with TMM. Among these, 80% of the samples are used as a training set, 10% for validation, and the remaining 10% for final testing. The training set is used for knowledge acquisition and training the feed forward networks; the validation set serves as a check to avoid overfitting, and the testing set evaluates the performance of the network.

## A. Interplay of reflection and transmission in non-Hermitian structures

Conservation laws, as fundamental physical principles, have been conventionally derived in the model-driven way and more recently re-discovered with data-driven approaches[42,43]. Typically, the elements of the transfer matrix result in a conservation relation that connects transmission and reflection for multilayer configuration. In Hermitian systems, the generalized conservation relation is simply expressed as $T + R = 1$ where left and right reflections are necessarily identical due to mirror symmetry i.e., $R_L = R_R = R$. Following this relation, Parity-time (*PT*)-symmetric systems with balanced gain and loss[44] hold the relation $\sqrt{R_L R_R} = |1 - T|$. In non-Hermitian systems, energy conservation is not valid due to the existence of gain and loss and, therefore, the intrinsic correlation between transmission and reflection in a general non-conservative system is yet to be developed. Here, we exploit deep learning to unveil the relation between transmission and reflection responses of non-Hermitian structures. We uncover one-to-one mapping from reflection $(R_L, R_R)$ to transmission $T$ (i.e., unique transmission response exists for any given reflection of arbitrary structure) and many-to-one in the reverse direction. The one-to-one relation is valuable to reconstruct the transmission from given reflections and analytically accessible through the training process that adjusts the weights in a neural network (NN) forming a series of nested ReLu functions. The trained NN provides a universal function approximator $f(\mathcal{R}, T, \Theta)$ where the function $f$ maps the reflection $\mathcal{R}$ to transmission $T$ with the trained weight parameters $\Theta$.

We implement the function $f$ by a fully connected neural network to learn the mapping between the left reflection, $R_L$ (or right reflection, $R_R$) and transmission, where the reflection is fed as input and transmission as output to the network [See Fig. 2a]. The training process optimizes weights $\Theta$ by minimizing the mean absolute loss $\mathcal{L} = \frac{1}{N}\sum_i |T_i - \hat{T}_i|$, where $T_i$ and $\hat{T}_i$ are the ground truth and the predicted transmission responses, respectively. The architecture of the designed network consists of four fully connected layers with 500–500–400–300 neurons activated by ReLu as depicted in Fig. 2a [see Supplementary information (SI) for network hyper parameters]. The performance of the trained network is quantified with the relative spectral error defined as: $e_s = \sum_i |T_i - \hat{T}_i|/T_i$, where $T_i$ and $\hat{T}_i$ denote the discretized target and predicted spectral response, respectively. In Fig. 3b, we show the distribution of the spectral error over the entire testing set in histogram form that indicates a high prediction accuracy (over 97%) of the network with an average error of 2.89% (dashed red line). The prediction performance of the trained model is demonstrated by three representative examples in Fig. 3 c-e, where the red and the black dashed curves corresponding to the TMM simulation and the DNN prediction, respectively, show an excellent agreement.

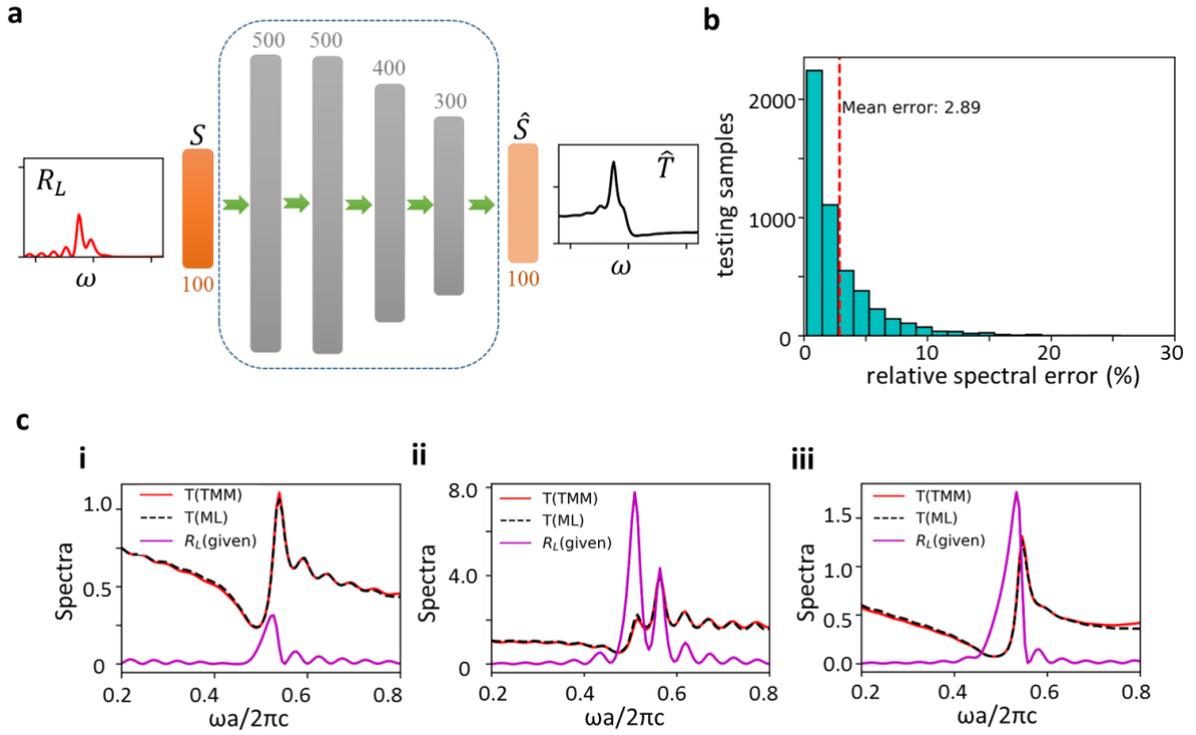

Fig. 2 Designed neural network for left reflection to transmission mapping. a Architecture of optimized network. b Histogram of spectral prediction error. c Representative examples for generation of transmission response from given reflection spectrum. The solid red and dotted black line represent the transmission calculated from TMM and ML method, respectively.

Note that neural networks can also be designed to determine the transmission spectra from the given right reflection, or from both reflections. The details are provided in SI. Our findings suggest that only one component of the reflection, either left or right, is sufficient to build the transmission profile of the considered non-Hermitian system.

**B. Sub-manifold learning for features recognition in non-Hermitian structures**

The spectral response of a photonic structure depends on a great extent on the operational frequency. A multi-dimensional frequency response presents a large number of features, which grows very fast with the number of discrete frequencies. Due to extra degrees of freedom introduced by gain-loss modulation in non-Hermitian systems, the spectral response of photonic structures becomes more complex. Consequently, it becomes extremely challenging to recognize the non-Hermitian structure from the spectral response and infer some knowledge for deep physical insight. Therefore, effective procedures to extract the most relevant information contained in the system response are highly desired. Different approaches have proven to be useful for extraction of the valuable information in data, such as PCA[40], multidimensional binary search tree[41], neural autoencoder[45], to name a few[46,47]. Such approaches are based on dimensionality reduction to visualize patterns, similarities and differences in data with minimum loss of information. In this study, we exploit the "*unsupervised*" dimensionality reduction algorithm using PCA[47] that transforms the discretized transmission spectra from a high-dimensional space to a low-dimensional latent space and models the sub-manifolds for different non-Hermitian structures. These manifolds, in the form of convex hull, are used to investigate the feasibility of having a desired optical response from a certain class of non-Hermitian structure. When mapped into the respective 2D feature space, these responses form a distribution of points, which characterize the correlation among the multi-frequency responses. The latent representation exploits the variance of features as well as the covariance between features to find major trends in spectral data. PCA organizes the large dimensions of each spectrum in terms of a respective feature vector, existing in higher-dimensional feature space. The procedure includes the rotation of the coordinate axes of the feature space such that the first axis results with the maximum possible data dispersion (as quantified by the statistical variance), the second axis with the second maximum dispersion, and so on[40]. This principle is illustrated for a transmission spectrum with the transmission amplitude at three different frequencies in Fig.1. In practice, we deal with $n$ frequencies and represent one transmission spectrum as an $n$-dimensional vector. Since the $n$ frequency features are not independent, a large set of the transmission spectra when all represented as $n$-dimensional vectors live on a low-dimensional manifold embedded in the $n$-dimensional space. Therefore, PCA can be employed to reduce the $n$-dimensional space to a lower-dimensional space (e.g., a 2D space spanned by the first two principal components), and visualize interesting patterns of gain, lossy, and balanced media [see Method section for details]. In our study, PCA is applied over the training data of transmission spectra and the first two components encoding the latent representation in 2D are shown in Fig. 3a. A convex hull is plotted to show the boundary of possible feasible response from the given non-Hermitian structure [see Fig. 3b]. Note that the convex hull has been formed with Quickhull algorithm to bound the transmission patterns in 2D latent space[48]. We identify three distinct manifolds corresponding to the effective gain, lossy, and balanced gain-loss class for the latent spectra. An overlapping area appears when the relative amount of gain and loss in the system are nearly equal. Around the center of the reduced space, the gain and loss become exactly balanced, which maximizes the unidirectionality. The difference in the area of color-coded sub-manifold implies the capability of the multilayer configuration to construct a wide range of feasible spectral responses with different non-Hermitian media. Sub-manifold learning can be used to forecast the feasibility of a response using a specific structure class.

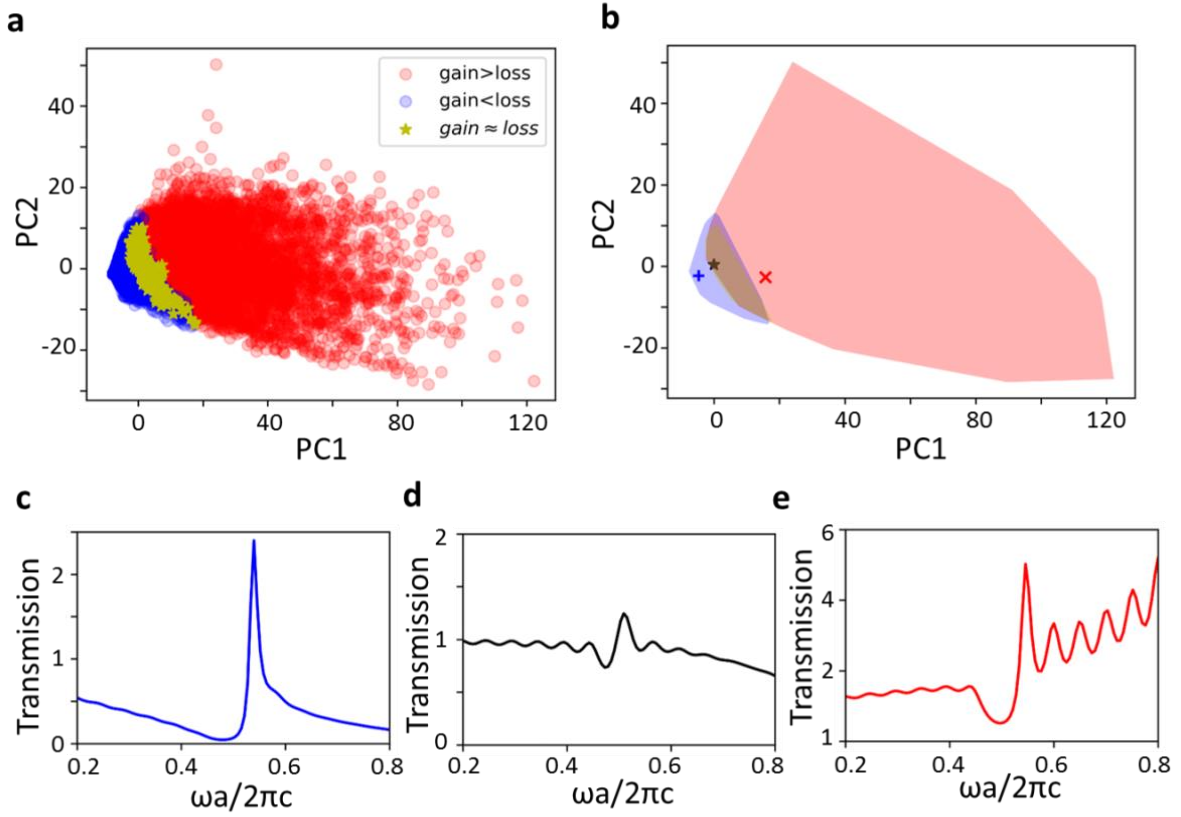

Fig. 3 Principal component analysis for the transmission spectra generated from a five-layer periodic non-Hermitian structure. a Representation of the first two principal components of the transmission spectra. b The corresponding convex-hulls of the feasible regions for non-Hermitian structures in Fig. 1 in the reduced 2D space. The red, blue, and yellow regions correspond to effective gain, lossy, and balanced media, respectively. The response generated from effective gain and lossy media resides top-right and bottom-left from the center of the latent space, respectively. An overlapping region is found when the gain and loss are balanced with the threshold $10^{-6}$. Within this overlapping area, PT-symmetry with perfect transmission exists around the center of the latent space. c Representative examples for the transmission spectra (and associated asymmetric reflections) residing in different region of reduced space are presented.

## C. Forward Neural Network

We develop a forward neural network to determine the spectral response for a given structure. It takes the design parameters as input $\mathcal{D}$ and the spectral response as ouput $S$ to build a mapping relation as $S = f(D)$ where $f$ is the complex nonlinear function constructed by the forward neural network. One-to-one mapping from design parameter to spectral response is a regression problem that can be modeled by NN, as depicted in Fig. 4a. The network is trained with mean absolute spectral loss defined as $\mathcal{L} = \frac{1}{N}\sum_i |S_i - \hat{S}_i|$ where $S_i$ and $\hat{S}_i$ are the ground truth and the predicted spectral response, respectively. The designed architecture for the forward NN has $500 - 500 - 400 - 400 - 300 - 300$ neurons in six layers, for transmission, and asymmetric reflections [see Fig.4a]. The details of training process and network hyper parameters are provided in SI. In order to assess the performance of the designed network, we define the relative spectral error on the testing data sets as: $e = \sum_i |S_i - \hat{S}_i|/S_i$ where $S_i$ and $\hat{S}_i$ correspond to target and predicted spectral responses. The average spectral errors (below 4%) for transmission and left/right reflection of the designed forward networks are presented in Fig. 4b [see the SI for more details]. The results of representative predicted responses agree well with TMM simulations, as show in Fig.4c,d.

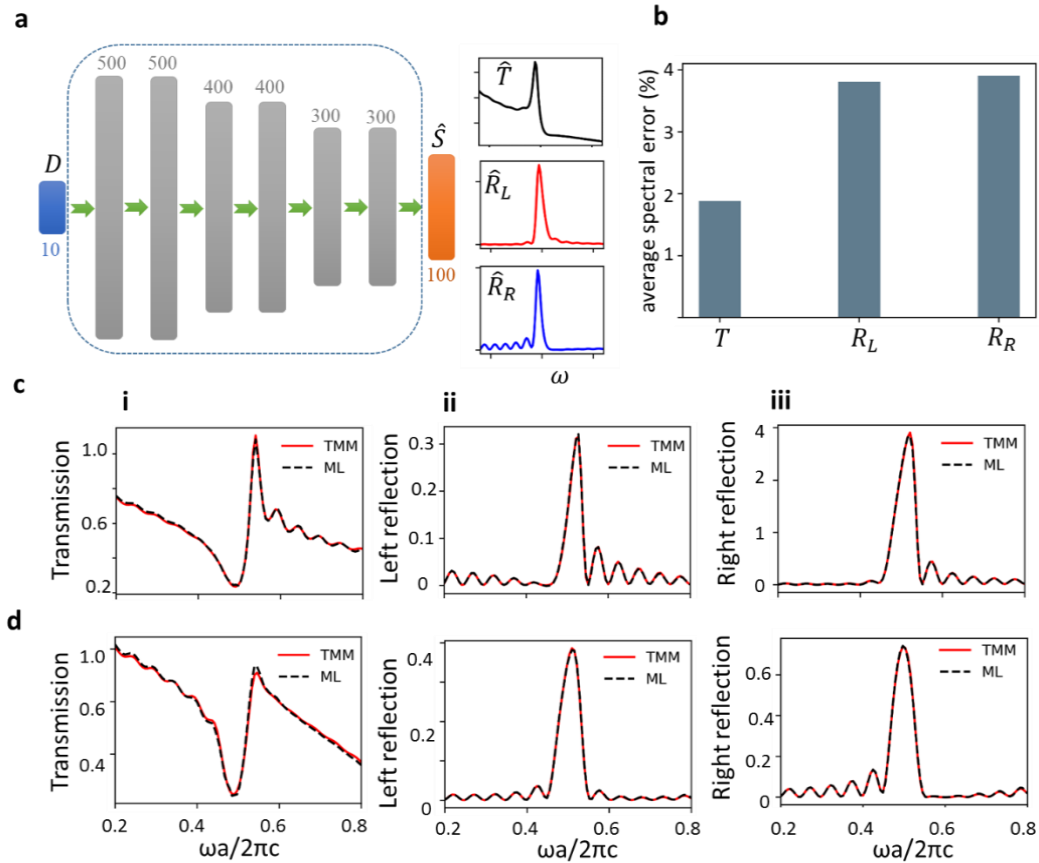

Fig. 4 Designed feed forward neural network for non-Hermitian structures. a Architecture of the 6-layer neural network. b Average spectral prediction error for transmission and left/right reflection. c-d Representative examples

of the predicted spectral response for the designed networks (i) transmission, (ii) left reflection, and (iii) right reflection.

### D. Inverse Network

Next, we move toward the inverse design problem after obtaining a well-trained forward NN model. Naturally, most inverse design problems are ambiguous due to the "one-to-many" mapping i.e., different non-unique solutions exist for the same target response. Consequently, a single discriminative network cannot achieve the optimal solution in the inverse design. To mitigate this problem, auxiliary training approaches and optimization strategies have been incorporated in the inverse design process[19,20,25,26,31,49-52]. In our work, we propose the sub-manifold learning with neural adjoint method to solve the inverse problem. To design a structure that results in a desired spectral response, the first step is to find the corresponding target point in the latent space by reducing the dimensionality of the desired response using PCA. Next, the KD-tree algorithm is applied to find the nearest neighboring point for the corresponding target point in the latent space within feasible sub-manifolds. The identified nearest neighboring point will act as the starting point for inverse design using a neural adjoint (NA) approach. The NA method determines the optimal solution by computing the gradient of the pretrained forward model $\hat{f}(D)$ with respect to the design parameters while keeping all weights and biases of the network fixed. The pretrained forward network $\hat{f}(D)$ provides a closed-form differentiable expression, and thus the calculation of $\partial \hat{f}/\partial D$ is trivial for the inverse design process. The gradient of the input design parameters with respect to a loss function $\mathcal{L}$ is estimated to iteratively move along the loss surface for the optimal solution. The inverse model can be denoted as $\hat{f}^{-1}(S, D_0)$, where $D_0$ is the initial structure obtained from sub-manifold learning and $S$ is the desired spectra.

Let $S$ be our desired spectra, and $D_i$ be our current best estimate of the design space, where the index $i$ indicates the iteration for adaptive gradient-based estimation procedure. To compute $D_{i+1}$, first we calculate the moving averages based on decaying exponential rates and gradient of the input design parameters with respect to a loss function in the following way:

$$m_i = \beta_1 m_{i-1} + (1-\beta_1)\frac{\partial \mathcal{L}(\hat{f}(\widehat{D}_i),S)}{\partial \widehat{D}_i}, \quad v_i = \beta_2 v_{i-1} + (1-\beta_2)\left[\frac{\partial \mathcal{L}(\hat{f}(\widehat{D}_i),S)}{\partial \widehat{D}_i}\right]^2, \quad (1)$$

where $\mathcal{L} = \frac{1}{N}\sum_i^N (\hat{f}(\widehat{D}) - S_i)^2$ is the mean squared loss function and $\beta_1, \beta_2$ are the exponential decay of the rates for the first moment estimates and second-moment estimates, respectively. Next, we need to correct the bias in the first and the second moment estimates as:

$$\widehat{m}_i = \frac{m_i}{1-\beta_1^i}, \quad \hat{v}_i = \frac{v_i}{1-\beta_2^i}. \quad (2)$$

Finally, we can update the design parameters based on the calculated moving averages with a step size $\alpha$:

$$\widehat{D}_{i+1} = \widehat{D}_i - \alpha \frac{\widehat{m}_i}{\sqrt{\hat{v}_i}+\epsilon}, \quad (3)$$

where $\alpha$ is the adaptive learning rate[53]. The major problem with existing neural adjoint methods is the selection of initial seed, however, we mitigate it with sub-manifold learning that eventually leads to an accurate solution in inverse design.

The results for the inverse design are summarized in Fig. 5. We consider three represented examples of desired responses whose transmission spectra resides within sub-manifold of lossy, gain, and balanced media indicated by b-d shown in Fig. 5a. Finding the nearest neighbor for the corresponding points in the latent space works as the initial structure that is evolved to achieve optimal target spectrum. The predicted spectra coincide with the target that indicates the strong capability of our method to design any spectral response absent in the training set, as shown in Fig. 5b-d. Figure 5c shows the perfect transmission and asymmetric reflection that do not require any strict space and time symmetry conditions as in the case of *PT*-symmetric structures. Therefore, the proposed approach can be used to design a more general class of planar reflectionless structures with non-Hermitian materials that can be realized with locally isotropic and non-magnetic materials.

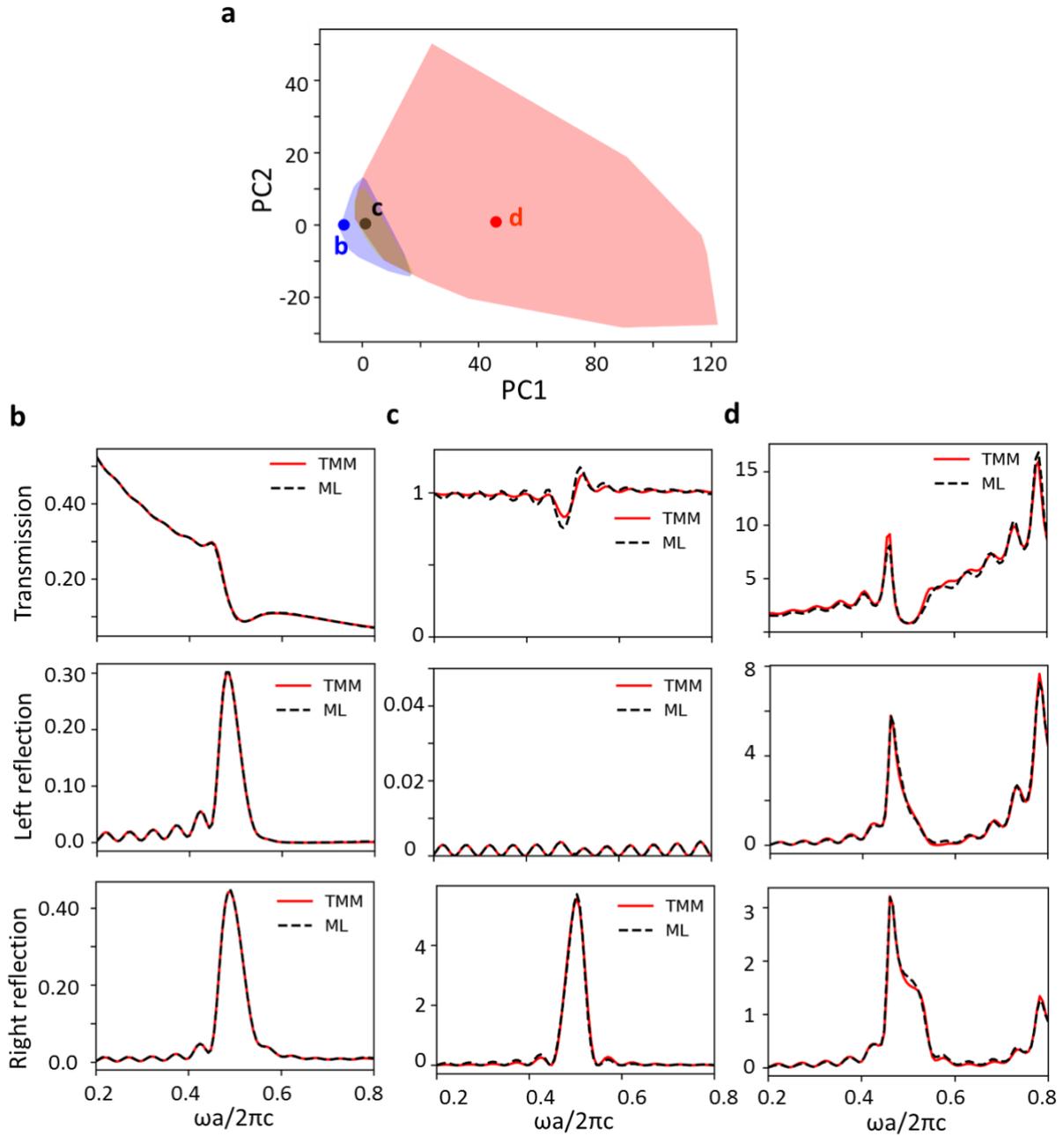

Fig. 5 Inverse design approach based on sub-manifold learning and neural adjoint method. a Convex-hulls of the feasible regions for non-Hermitian structures for the training transmission data in latent space. b-d Desired spectral responses designed with adaptive gradient descent method. The initial seed is obtained within feasible sub-manifolds of lossy, gain, and balanced non-Hermitian systems.

**Conclusion**: We propose unsupervised and supervised learning for modeling the optical response of non-Hermitian photonic structures that can derive valuable insights about the relationships of asymmetric reflections and transmission in non-conservative systems. The study reveals that either the left or the right reflection is sufficient to determine the corresponding transmission profile. As demonstration, we study a multilayered periodic structure and develop a machine learning framework for knowledge acquisition and retrieval of the design parameters for a given spectral response. In particular, we propose a machine learning model to determine the transmission from a given asymmetric reflection without design parameters that uncover the one-to-one mapping from the reflection to the transmission spectra. The dimensionality reduction based on PCA is applied to learn the sub-manifolds of lossy, gain, and balanced structures via the transmission response. The learned sub-manifolds in 2D latent space are useful to determine the feasibility of the response with a given class of structures and find the best initial seed for inverse-design with neural adaptive gradient method. The approach is used to inversely design the unidirectional properties with multilayered structures that do not require any strict symmetry as in *PT*-symmetric structures. Our methodology integrates dimensionality reduction to automate the design and, thus, provides a versatile platform to learn new physical insights in non-Hermitian photonics structures. The inverse design method is not restricted to optical problems, and it can be applied directly to find accurate solutions to ill-posed problems in other physical systems such as acoustics, elasticity, or plasmonics.

## Methods

**Transfer Matrix Method (TMM):**

TMM is one of the most widely used methods to model light propagation in multilayered structures. In this study, we consider a one-dimensional periodic structure formed by five alternating layers with coherent interfaces. We assume that all layers are isotropic and nonmagnetic with different dielectric material. The electric field is represented as a superposition of the left- and right- propagating waves with wavevector, $k=\omega/c$, as: $E^-(z) = E_L^+ e^{ikz} + E_R^- e^{-ikz}$ and $E^+(z) = E_R^- e^{ikz} + E_R^+ e^{-ikz}$ for the left $(z < -L/2)$ and the right $(z > L/2)$ side of the structure, respectively. The continuity conditions for the tangential electric field components at the outer interfaces of the layered structure determine the transfer matrix, $M$, that is the product of matching and propagation matrices. Mathematically, the transfer matrix relates the field amplitudes of the left- and right propagating waves in the following manner:

$$\begin{pmatrix} E_R^- \\ E_R^+ \end{pmatrix} = M \begin{pmatrix} E_L^+ \\ E_L^- \end{pmatrix}, \quad M = \begin{pmatrix} M_{11} & M_{12} \\ M_{21} & M_{22} \end{pmatrix}. \tag{4}$$

The transmission coefficient $t_{R,L}$, and reflection coefficient $r_{R,L}$, (along with the transmittance $T_{R,L} = |t_{R,L}|^2$ and reflectance $R_{R,L} = |r_{R,L}|^2$) for left (L) and right (R) incidence waves can be computed from boundary conditions and related with the elements of transfer matrix as:

$$t_R = \frac{1}{M_{22}}, \quad t_L = \frac{M_{11}M_{22} - M_{12}M_{21}}{M_{22}}$$

$$r_R = \frac{M_{12}}{M_{22}}, \quad r_L = -\frac{M_{21}}{M_{22}}$$

(5)

**Principal Component Analysis:**

PCA is a powerful unsupervised method for dimensionality reduction that transforms the data to a lower dimensional space to identify the intrinsic patterns and correlation in the data without loss of original information. Consider *m* data points of *n*-dimensional spectral space S = [$s_1$, $s_2$..... $s_n$] where, $s_1$ represents $i^{th}$ features, and *n* represents the total number of features. The response data matrix can be written as $R^{m \times n}$ from which successive *k* orthogonal components (also called principal components) are computed to find the direction of the maximum variance. The largest eigenvalues of the response data matrix along with the corresponding eigenvectors are used to analyze a large amount of high-dimensional transmission data. In this work, we implement PCA with python sklearn libaray that uses the singular value decomposition for the calculation of principal components.

## Acknowledgements

The work described in here is supported by King Abdullah University of Science and Technology (KAUST) Artificial Intelligence Initiative Fund and KAUST Baseline Research Fund No. BAS/1/1626-01-01. K.S. acknowledges funding from European Social Fund (project No 09.3.3- LMT-K712-17- 0016) under grant agreement with the Research Council of Lithuania (LMTLT).